\documentclass[12pt,preprint]{aastex}

\begin{document}
\title{An Updated Search of Steady TeV $\gamma-$Ray Point Sources
in Northern Hemisphere Using the Tibet Air Shower Array}
%\author{
%(The Tibet AS$\gamma$ Collaboration)}
\author{
Y.~Wang\altaffilmark{1}
  ~~X.~J.~Bi\altaffilmark{1}
 ~~S.~W.~Cui\altaffilmark{2}
  ~~L.~K.~Ding\altaffilmark{1}
 ~~Danzengluobu\altaffilmark{3}
 ~~X.~H.~Ding\altaffilmark{3}
~~C.~Fan\altaffilmark{4}
~~C.~F.~Feng\altaffilmark{4} 
 ~~Zhaoyang Feng\altaffilmark{1}
~~Z.~Y.~Feng\altaffilmark{5}
 ~~X.~Y.~Gao\altaffilmark{6}
~~Q.~X.~Geng\altaffilmark{6}
 ~~H.~W.~Guo\altaffilmark{3}
~~H.~H.~He\altaffilmark{1}
 ~~M.~He\altaffilmark{4}
 ~~Haibing~Hu\altaffilmark{3} 
 ~~H.~B.~Hu\altaffilmark{1}
~~Q.~Huang\altaffilmark{5}
 ~~H.~Y.~Jia\altaffilmark{5}
~~Labaciren\altaffilmark{3}
 ~~G.~M.~Le\altaffilmark{7}
 ~~A.~F.~Li\altaffilmark{4}
 ~~J.~Y.~Li\altaffilmark{4}
~~Y.-Q.~Lou\altaffilmark{8} 
 ~~H.~Lu\altaffilmark{1}
 ~~S.~L.~Lu\altaffilmark{1}
 ~~X.~R.~Meng\altaffilmark{3}
 ~~J.~Mu\altaffilmark{6}
 ~~J.~R.~Ren\altaffilmark{1}
 ~~Y.~H.~Tan\altaffilmark{1}
~~B.~Wang\altaffilmark{6}
 ~~H.~Wang\altaffilmark{1}  
~~Y.~G.~Wang\altaffilmark{4}
 ~~H.~R.~Wu\altaffilmark{1}
 ~~L.~Xue\altaffilmark{4}
 ~~X.~C.~Yang\altaffilmark{6}
  ~~Z.~H.~Ye\altaffilmark{9}
~~G.~C.~Yu\altaffilmark{5}
 ~~A.~F.~Yuan\altaffilmark{3}
  ~~H.~M.~Zhang\altaffilmark{1}  
 ~~J.~L.~Zhang\altaffilmark{1}
 ~~N.~J.~Zhang\altaffilmark{4}
 ~~X.~Y.~Zhang\altaffilmark{4}
~~Y.~Zhang\altaffilmark{1}
 ~~Yi~Zhang\altaffilmark{1}
~~Zhaxisangzhu\altaffilmark{3}
 ~~X.~X.~Zhou\altaffilmark{5} and ~~Q. Yuan\altaffilmark{1}
}

\altaffiltext{1}{Key Laboratory of Particle Astrophysics, Institute of High Energy Physics,
   Chinese Academy of Sciences, Beijing 100049, China.}
\altaffiltext{2}{Department of Physics, Hebei Normal University, Shijiazhuang 050016, China.}
\altaffiltext{3}{Department of Mathematics and Physics, Tibet University, Lhasa 850000, China.}
\altaffiltext{4}{Department of Physics, Shandong University, Jinan 250100, China.}
\altaffiltext{5}{Institute of Modern Physics, Southwest Jiaotong University, Chengdu 610031, China.}
\altaffiltext{6}{Department of Physics, Yunnan University, Kunming 650091, China.}
\altaffiltext{7}{National Center for Space Weather, China Meteorological
Administration, Beijing 100081, China.}
\altaffiltext{8}{Physics Department and Tsinghua Center for Astrophysics, Tsinghua University, Beijing 100084, China.}
\altaffiltext{9}{Center of Space Science and Application Research, Chinese Academy of Sciences, Beijing 100080, China.}

\begin{abstract}
Using the data taken from Tibet II High Density (HD) Array (1997
February-1999 September) and Tibet-III array (1999 November-2005
November), our previous northern sky survey for TeV $\gamma-$ray
point sources has now been updated by a factor of 2.8 improved
statistics. From $0.0^{\circ}$ to $60.0^{\circ}$ in declination
(Dec) range, no new TeV $\gamma-$ray point sources with
sufficiently high significance were identified while the
well-known Crab Nebula and Mrk421 remain to be the brightest TeV
$\gamma-$ray sources within the field of view of the Tibet air
shower array. Based on the currently available data and at the
90\% confidence level (C.L.), the flux upper limits for different
power law index assumption are re-derived, which are approximately
improved by 1.7 times as compared with our previous reported limits.
\end{abstract}

\keywords{AS$\gamma$ experiment, $\gamma-$ray point sources, 90\% C.L., flux
upper limits}

\section{Introduction}
The development of TeV $\gamma-$ray observations has experienced a
revolutionary progress$^{[1,2]}$ since the finish of our northern
sky survey work$^{[3]}$ (hereafter Paper I). For 
example, High Energy Stereoscopic System (HESS)
experiment alone has discovered more than 40 new $\gamma-$ray
sources in southern hemisphere with unprecedented angular
resolution and sensitivity. Together with other sensitive Imaging
Air Cherenkov Telescopes (IACTs), such as Major Atmospheric Gamma
Imaging Cherenkov Telescope (MAGIC), Collaboration of Australia
and Nippon (Japan) for a GAmma Ray Observatory in the Outback (CANGAROO), Very
Energetic Radiation Imaging Telescope Array System (VERITAS), more
than 50 new sources have been discovered in the past several years
and the number as well as the diversity of TeV $\gamma-$ray
sources have been increasing. Spatial and temporal information of
these sources are now available with high accuracy and this makes
it possible for further studies on acceleration of high-energy
cosmic rays, relativistic astrophysics, as well as quantum gravity
theory and so forth. To demonstrate the advantage of its wide
field of view and high duty cycle, the Tibet Air Shower Array
experiment performed a high-precision measurement on the
two-dimensional (2D) anisotropy of cosmic rays in the energy range
of a few to several hundred TeV and discovered a fairly compact
new anisotropic component in the direction of Cygnus
region$^{[4]}$. Furthermore, MILAGRO experiment has discovered an
extended $\gamma-$ray source in the direction of Cygnus
region$^{[5]}$ and a few more other sources in the Galactic
plane$^{[6]}$, in addition to the diffuse $\gamma-$ray emission
from the Galactic plane$^{[7]}$. While some of the MILAGRO sources
were confirmed or supported by the Tibet Air Shower
experiment$^{[4, 8, 9]}$, it would be extremely interesting and
important for the Tibet Air Shower experiment to systematically
update its northern sky survey with a much larger data sample
currently available.

\section{Tibet Air Shower Array Experiment and Observations}
The Tibet air shower array experiment has been successfully
carried out at Yangbajing Cosmic Ray Station ($90.522^{\circ} $E,
$\ 30.102^{\circ} $N) in Tibet, China, since 1990, at an altitude
of $4300$m above sea level. Having been upgraded several
times$^{[10,11,12]}$,
the Tibet HD and III arrays have identical structures except
the array size and shape. A 0.5 cm thick lead plate was later
placed on top of each counter to improve fast-timing (FT) data by
converting $\gamma$ rays into electron-positron pairs. The angular
resolution was first estimated from Monte Carlo (MC) simulations
and then confirmed experimentally by observing the Moon shadow to
be about $0.9^\circ$ in the energy range above 3 TeV.
The data used in this analysis were collected by running the Tibet
HD array for 555.9 live days from 1997 February to 1999 September
and the Tibet III array for 1318.9 live days from 1999 November to
2005 November. The events are selected by imposing five criteria
on the reconstructed data: (1) Each shower event should fire four
or more FT detectors recording 1.25 or more particles. (2) The
estimated shower center location should be inside the detector
array. (3) $\sum\rho_{FT}$ should be larger than 15, where
$\sum\rho_{FT}$ is the sum of the number of particles per square
meter detected in each detector. (4) The zenith angle of the
incident direction should be smaller than $40^{\circ}$. (5) The
residual error in direction reconstruction should be less than
1.0m. After applying these cuts and a data quality controll, about
$2.0\times10^{10}$ shower events were available for our data
analysis here.

\section{Data Analysis}
Based on the successful analysis of Paper I and
for simplicity, Method II (i.e., the all-distance ``equi-zenith
angle'' method) was adopted to construct the 2D cosmic ray
intensity map with pixels in the size of
$0.1^{\circ}\times0.1^{\circ}$ in equatorial coordinate.
The idea of this method is that at any moment, for all directions,
if we scale down (or up) the number of observed events by dividing
them by their relative cosmic ray intensity, then those scaled
numbers of events in a zenith angle belt should be equal
anywhere in the sense of statistics. A $\chi^2$ function can be built
accordingly, the relative intensity of cosmic rays $I(R.A.,Dec)$ and
its error $\Delta I(R.A.,Dec)$ in each direction can be solved by 
minimizing the $\chi^2$ function. It is worth mentioning that
source information in a $0.1^{\circ}\times0.1^{\circ}$ bin had been 
included in the intensity $I(R.A.,Dec)$. For details of this 
method, the reader is referred to Paper I.
To remove the large-scale cosmic ray anisotropy and to keep
the local event excess structure which is due to the $\gamma-$ray
emission, we use the similar subtraction procedure as in Paper I
when parameterizing the projected intensity distribution along the
right ascension (R.A.) direction for any Dec belt. After subtracting the anisotropy, we can obtain the relative intensity of cosmic rays $I_{corr}(R.A.,Dec)$  and its error $\Delta I_{corr}(R.A.,Dec)$.
The number of excess events and their uncertainties in cell (R.A.,Dec) can be calculated as
\begin{equation}\label{equ1}
N_s(R.A.,Dec)=[I_{corr}(R.A.,Dec)-1]N_{obs}(R.A.,Dec)/I_{corr}(R.A.,Dec) 
\end{equation}
\begin{equation}\label{equ1}
\Delta N_s(R.A.,Dec)=\Delta I_{corr}(R.A.,Dec)N_{obs}(R.A.,Dec)/I_{corr}(R.A.,Dec) 
\end{equation}
where $N_{obs}(R.A.,Dec)$ is the number of events in an on-source bin.

Given the angular resolution of the Tibet air shower array,
events are summed up from a cone with an axis pointing to the source direction,
and the half-opening angle is set as $0.9^{\circ}$(for $E>3$TeV)$^{[13]}$ or $0.4^{\circ}$(for $E>10$TeV). All celestial cells with their centers located 
inside the cone contribute to the number of events as well as its uncertainty.
Finally, the significance for an on-source window centered at the cell
$(R.A._{on},Dec_{on})$ can be calculated by
\begin{equation}\label{equ2}
S(R.A._{on},Dec_{on})=\frac{\sum_{(R.A.,Dec)\in{cone}}\{[I_{corr}(R.A.,Dec)-1]
N_{obs}(R.A.,Dec)/I_{corr}(R.A.,Dec)\}}{\sqrt{\sum_{(R.A.,Dec)\in{cone}}
[\Delta I_{corr}(R.A.,Dec)N_{obs}(R.A.,Dec)/I_{corr}(R.A.,Dec)]}}
\end{equation}
The systematic uncertainty for the significance value due to the
above subtraction procedure on large-scale anisotropy is estimated
to be 0.2$\sigma$ by adjusting the bin size and the smoothing
parameters.

\section{Results and Conclusions }
 Distribution of significance for all bins in the surveyed sky
 are shown in Fig.1. It agrees very well with a normal distribution
 on the negative side, indicating that systematic effects are well
 under control. The positive side contains more high-significance
 entries than those expected from pure statistical fluctuations, and
 they are related to two well-known TeV $\gamma-$ray
 sources, namely the Crab Nabula and Mrk421. After removing their
 contributions, in such a way that those cells within $2^{\circ}$
 regions around the Crab Nebula and Mrk421 are excluded, we get the
 dash-dotted histogram as shown in Fig.1, consistent with the expectations
 from random background fluctuations.
\vspace{1.cm}
  \begin{figure}[htbp]
   \begin{center}
  \vspace{-0.3cm}
  \includegraphics[width=23pc,height=20pc]{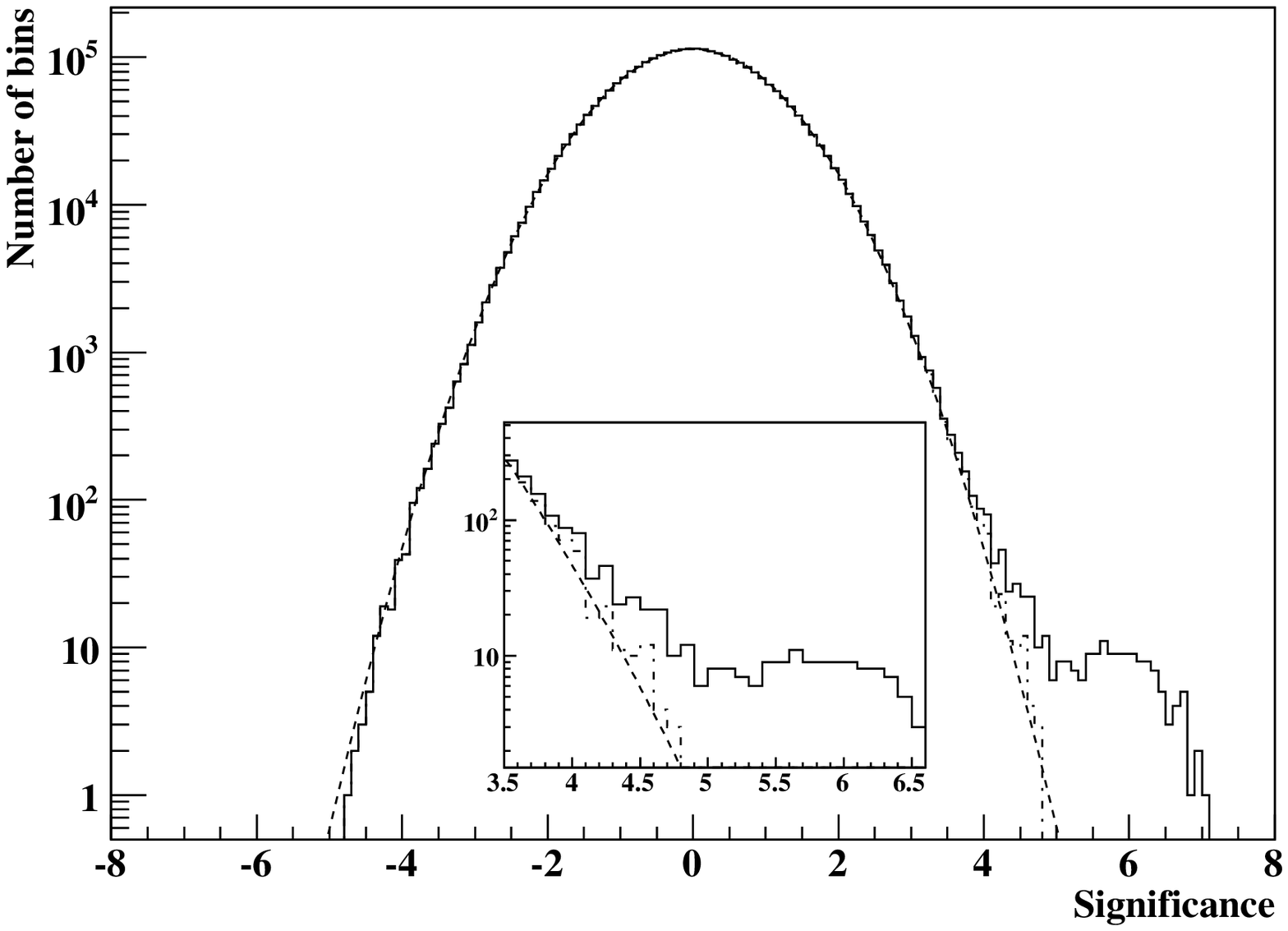}
\end{center}
\end{figure}
 \begin{center}
\vspace{-0.7cm}
\parbox[c]{14cm}{\footnotesize{\bf Fig.1}
The significance map is shown here. The solid curve is derived
from all cells defined in the analysis. The dash-dotted histogram
excludes cells close to the Crab Nebula or Mrk421. The dashed line
represents the best fit of a Gaussian curve to the data, its mean is $-0.002\pm0.01$ and standard deviation is $1.013\pm0.005$. }
\end{center}
\iffalse
\begin{figure}[htbp]
   \begin{center}
 \includegraphics[width=30pc,height=30pc]{allsky_1D_sig_tstv1_forgaoh.eps}
\end{center}
\caption{The significance map is shown here. The solid curve is derived
from all cells defined in the analysis. The dash-dotted histogram
excludes cells close to the Crab Nebula or Mrk421. The dashed line
represents the best fit of a Gaussian curve to the data, its mean is $-0.002\pm0.01$ and standard deviation is $1.013\pm0.005$.}
\end{figure}
\fi
For our updated sky survey, the information of five candidates for
possible $\gamma-$ray sources, each with an excess of greater than
4.5$\sigma$, are summarized in Table 1. Only the pixel with the
highest significance from each independent direction is listed.

%\eject\vfill

%%%%%%%%%%%%%%%%%%%%%%%%%%%%%%%%%%%%%%%%%%%%%%%%
 \begin{table}[!h]
       \begin{center}
{Table 1: Candidate locations of all directions with an excess greater than $4.5\sigma$ }\\
        \begin{tabular}{l c c c c c c c }\hline
   No. &   $R.A.$ & $Dec $ & $N_{ON}$ & $N_{OFF}$ & $N_{S}$ & $\bigtriangleup N_{S}$ & $S_{pretrials}$\\
1   & 57.95  & 53.25 &  2405072.8 &  2397926.7 &  7146.1 &  1548.5  & 4.6 \\
2   & 70.55 & 11.35 &  2306840.6 &  2299785.4 &  7055.2  & 1516.5 &  4.7 \\
$3^a$ & 83.75 &  21.95 &  3078848.1  & 3066434.9  & 12413.3 &  1751.1 &7.1 \\
4   & 89.45  & 30.05  & 3359526.5  & 3350799.7  & 8726.8 &  1830.5  & 4.8 \\
$5^b$ & 166.25 &  38.25 &  3301780.3  & 3292945.8 &  8834.4 &1814.6 &  4.9 \\
          \hline
          \end{tabular}
        \end{center}
%%%%%%%%%%%%%%%%%%%%%%%%%%%%%%%%%%%%%%%%%%%%%%%%%%%
 \vspace{-0.5cm}
 \begin{center}
\parbox[c]{14cm}{\footnotesize{\bf Table 1}
The columns are (from left to right) sequence of prominent
direction, R.A.(J2000), Dec(J2000), number of measured events in
on-source window ($N_{ON}$), number of background events
($N_{OFF}$), event number excess in on-source window (
$N_{S}$=$N_{ON}$-$N_{OFF}$), uncertainty on the event number excess
($\bigtriangleup N_{S}$), and the significance
$S_{pretrials}$ of deviation $N_{ON}$ from $N_{OFF}$.\\
{\bf Notes:} R.A. and Dec columns are due to the way we divide the
bin in the analyses; $3^a$-----The Crab Nebula and
$5^b$-----Mrk421.}
\end{center}
\end{table}
As can be seen in Table 1, the list includes two established
sources Crab Nebula and Mrk421 which remain to be the brightest
TeV $\gamma-$ray sources in the northern sky. Compared with Paper
I, the significance of the Crab Nebula is increased from
$5.0\sigma$ to $7.1\sigma$ at the highest significance position
(from $4.1\sigma$ to $6.0\sigma$ at the nominal position of the
Crab Nebula, which is consistent with the expected enhancement
from $3.7\sigma$ to $6.2\sigma$ within error bars according to the
changing statistics); while the significance of Mrk421 is dropped
somewhat mainly due to the fact that Mrk421 is not a stable
source; it happens to be in a high state with the data used in
Paper I but remains less active$^{[14]}$ in the succeeding period
when the data are used in the current analysis. As for the other
two candidates in Table 1, No.2 and No.4 have also appeared in Paper I;
however their significance values are slightly decreased after we
have included more data into this analysis.
There are still two other sources located at ($70.45^{\circ}$,\
$18.05^{\circ}$) and ($221.75^{\circ}$,\ $32.75^{\circ}$) which passed in
Paper I but failed this time, and the current analysis selects a new
candidate at position ($57.95^{\circ}$,\ $53.25^{\circ}$) with a
significance value just above $4.5\sigma$. It should be noted that
the above$-$discussed phenomena, except the Crab Nebula and
Mrk421, are probably due to the background fluctuations and also
possibly due to their intrinsic unstable features;
the conclusive results will rely on a further data analysis. In
summary, compared with Paper I, the number of hot spots reduces
from 4 to 3 (only for the candidates determined from Method II and
by excluding the two known sources: the Crab Nebula and Mrk421). 
Both are consistent with the expectation from statistic fluctuation. With 200 Toy MC experiments, the numbers of hot spots(each satisfies 4.5$\sigma$ requirement)
are obtained for each experiment, the probabilities to observe no less 4 and 3 are found to be 8\% and 26\% respectively.  In addition, the
locations of candidates are somewhat different between the two
observations. As for the difference, it agrees with the pointing
accuracy of the Tibet AS$\gamma$ array. %Taking a MC simulation for
%the Crab Nebula as an example, we estimate the probability to be
%40\% when the largest significance position difference is
%$0.4^{\circ}$ between two observations.
Taking Crab Nebula ($0.4^{\circ}$ position difference between two observations)
as an example, we estimate the probability with MC experiments and find 40\%
 of the MC experiments have a position difference no less than $0.4^{\circ}$.  
Nevertheless, given the large number of trials, the significance
values from all directions other than the Crab Nebula and Mrk421
are not high enough to definitely claim any existence of a new
point source, although they will become clearer with the future
improved statistics of observational data or can be interesting
regions for further follow-up observations with more sensitive
IACTs.

%It should be noted that
It is worth mentioning that the two MILAGRO newly reported TeV
sources, MGRO J1908+06$^{[6]}$ and MGRO J2019+37$^{[5]}$ not listed
in Table 1 due to their lower significance, had been our two candidates
with only marginal yet persistent event excess. In Paper I,
we found $4.8\sigma$ on ($286.65^{\circ},\ 5.55^{\circ}$),
$0.4^{\circ}$ angular separation from MGRO J1908+06. However the
significance value is $4.3\sigma$ in the current analysis, not
scaled up with the increase of statistics but consistent with the
expectation based on the flux measured by MILAGRO$^{[6]}$ and
HESS$^{[15]}$. Another interesting point source, close to an
extended source MGRO J2019+37, has been discussed several times in
Tibet AS$\gamma$ papers$^{[4,16,17]}$. Our dedicated analysis has
reported a preliminary $5.8\sigma$ excess in ($304^{\circ},\
36.1^{\circ}$)$^{[9]}$ and this result should be regarded as a
confirmation of the MILAGRO's discovery$^{[2]}$. While less
sensitive to such an extended source, the current point source search
analysis still finds a $4.0\sigma$ excess in the direction of
($303.25^{\circ},\ 35.95^{\circ}$).

Based on the above analysis we know that the significance from all
directions other than the Crab Nebula and Mrk421 are not high
enough to definitely claim any existence of a new point source, we
set a 90\% C.L. upper flux limit for all directions in the sky,
except at the positions of the Crab Nebula and Mrk421. The
prescription of Helene$^{[18]}$ is used to calculate the upper
limits of the number of signal events at the 90\% C.L. for
energies higher than 3TeV and 10TeV from each region of the
northern sky. Then the effective detection area of the Tibet air
shower array is evaluated by full MC simulation assuming a
Crab-like $\gamma-$ray spectrum $E^{-2.6}$ for a set of Dec values
($0.0^{\circ}$, $10.0^{\circ}$, $20.0^{\circ}$, $30.0^{\circ}$,
$40.0^{\circ}$, $50.0^{\circ}$, and $60.0^{\circ}$) and
interpolated to other Dec values between $0.0^{\circ}$ and
$60.0^{\circ}$. Taking into account the live time, the newly
derived 90\% C.L. average flux upper limit along the R.A.
direction as a function of Dec is shown in Fig. 2(a), which is
$(0.8\sim1.9)\times10^{-12} \hbox{ cm}^{-2}\hbox{ s}^{-1}$ for
$E>3$TeV and $(1.3\sim2.5)\times10^{-13} \hbox{ cm}^{-2}\hbox{
s}^{-1}$ for $E>10$TeV respectively.
Additionally, since the response of the Tibet air shower array is
energy dependent, the flux upper limits obtained from these data are
dependent on the energy spectra of the possible sources of TeV gamma
rays. The same procedure is applied to the cases of other power-law
indices for energy above 3TeV and 10TeV, the corresponding average
flux limits can be found in Fig.2(b), which are
$(0.8\sim2.2)\times10^{-12} \hbox{ cm}^{-2}\hbox{ s}^{-1}$ for
$E>3$TeV and $(1.2\sim3.0)\times10^{-13} \hbox{ cm}^{-2}\hbox{
s}^{-1}$ for $E>10$TeV respectively and have approximately 1.7
improvement compared with the reported limits in Paper I. These
limits are well consistent with the fact that the majority of
$\gamma-$ray sources discovered in recent years have integrated
fluxes less than 10\% of those of the Crab Nebula at 1TeV
energy$^{[19]}$.

  \vspace{1.cm}
  \begin{figure}[htbp]
  \vspace{-0.6cm}
  \includegraphics[width=0.49\textwidth]{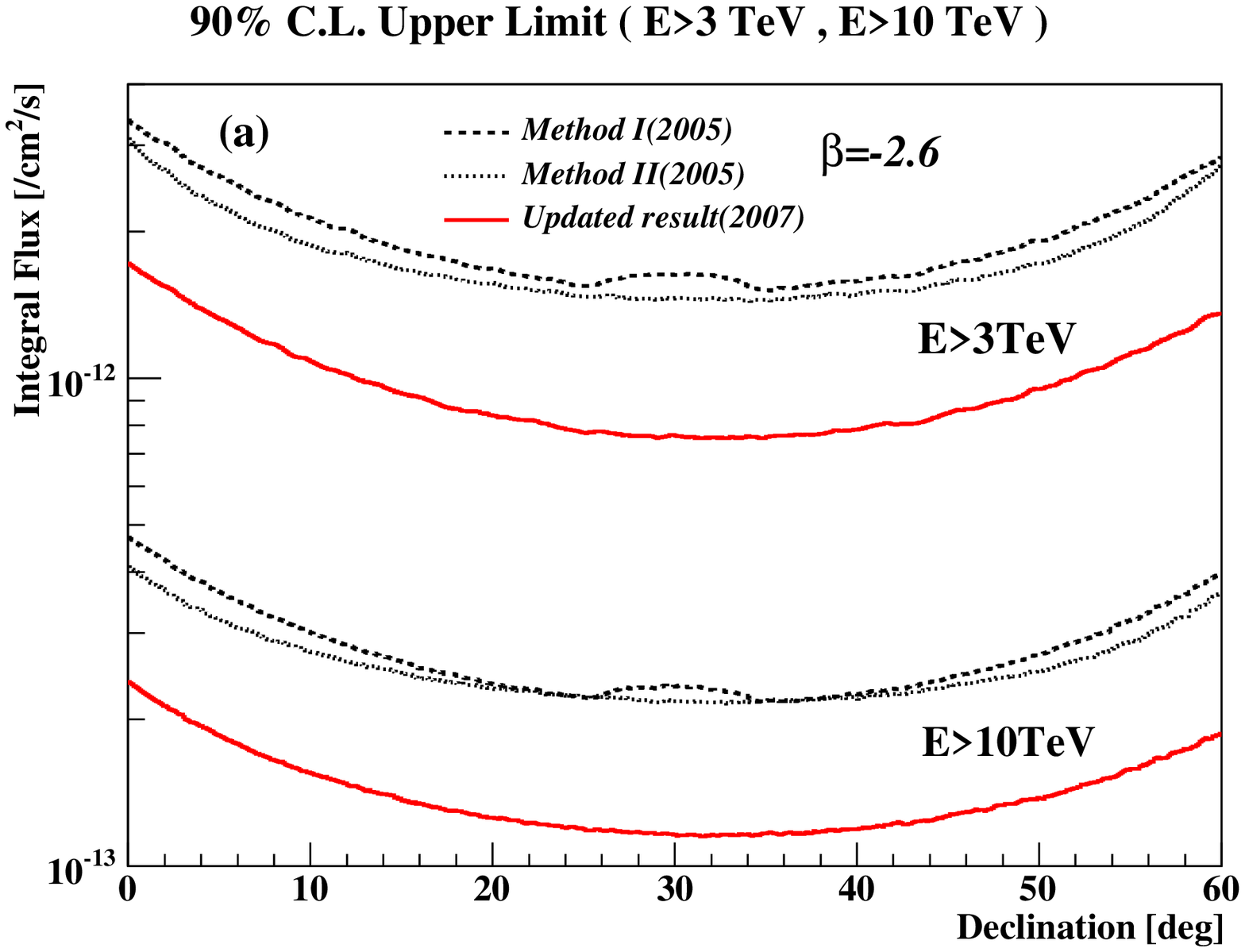}
  \hspace{0.5cm}
\includegraphics[width=0.52\textwidth]{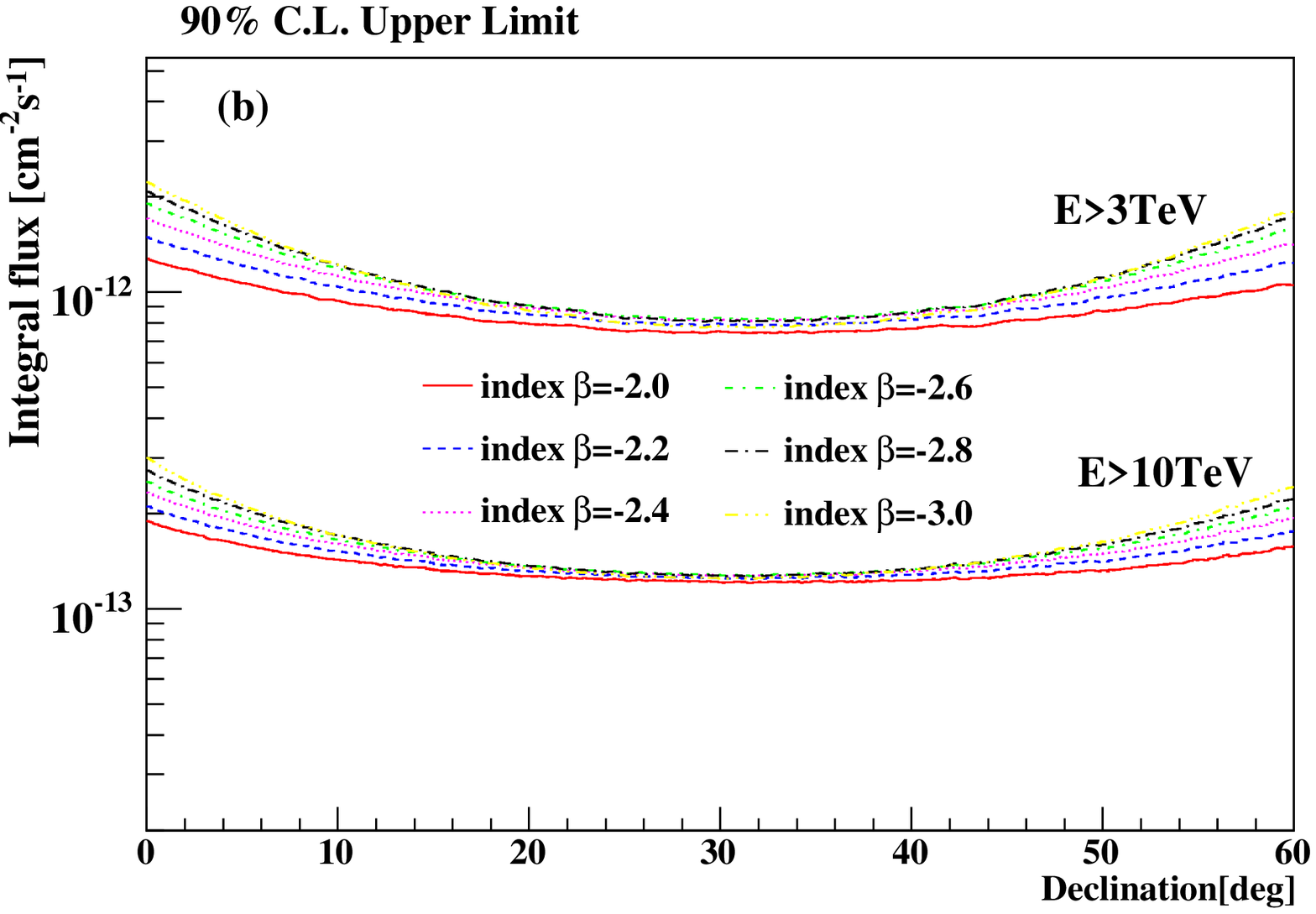}
   \vspace{-0.5cm}
 \begin{center}
\parbox[c]{13cm}{\footnotesize{\bf Fig.2}
R.A. direction-averaged 90\% C.L. upper limit on the integral flux
above 3TeV and 10TeV. {\bf (a)} for a Crab-like point source, i.e.,
with an energy spectrum of $E^{-2.6}$; {\bf (b)} for different indices
of power-law spectra.
}
\end{center}
  \end{figure}
In conclusion, we performed an updated northern sky survey for the TeV $\gamma-$ray
point sources in a Dec band between $0.0^{\circ}$ and $60.0^{\circ}$ using about
eight-year data obtained from February 1997 to November 2005 by the Tibet air shower
array. The significance except Crab and Mrk421 is not high enough to definitely
claim any existence of new sources. Accordingly, more stringent 90\% C.L. flux upper
limits than the ones in Paper I are set from the rest of positions based on the
assumption that candidate point sources have power-law spectra with indices varying
from 2.0 to 3.0. In the near future, we will add a large muon detector array under
the Tibet air shower array for the purpose of increasing its
$\gamma-$ray sensitivity in the 100 TeV energy region (10-1000
TeV) by discriminating between $\gamma$ rays and the cosmic$-$ray
hadrons$^{[20]}$. According to a full MC simulation, flux
sensitivity of this new project will be an order or more better
than the present one in the 100TeV region$^{[21]}$. Approximately
10 new sources are expected to be discovered and we will be able
to measure the cutoff energies of known and unknown sources which
are potential origins of Galactic cosmic rays.

\acknowledgments{}The collaborative experiment of the Tibet Air
Shower Arrays has been performed under the auspices of the Ministry
of Science and Technology of China and the Ministry of Foreign
Affairs of Japan. This work is supported in part by Grants-in-Aid
for Scientific Research on Priority Areas (712) (MEXT), by the Japan
Society for the Promotion of Science (JSPS), by the National Natural
Science Foundation of China (in part by NSFC Grants 10675134 and
10533020) and by the Chinese Academy of Sciences.

%%%%%%%%%%%%%%%%%%%%%%%%%%%%%%%%%%%%%%%%%%%%%%%%%%%%%%%%%
%\end{multicols}

%\begin{multicols}{2}
%%²Î¿¼ÎÄÏ×

\end{document}